\documentclass{PoS}

\usepackage{epsfig}
\usepackage{latexsym}
\usepackage{graphicx}
\usepackage{bm}
\usepackage{longtable}

\def\lsim{\raise0.3ex\hbox{$<$\kern-0.75em\raise-1.1ex\hbox{$\sim$}}}
\def\gsim{\raise0.3ex\hbox{$>$\kern-0.75em\raise-1.1ex\hbox{$\sim$}}}
\def\simgt{\rlap{\lower 3.5 pt\hbox{$\mathchar \sim$}}\raise 1.0pt \hbox {$>$}}
\def\simlt{\rlap{\lower 3.5 pt\hbox{$\mathchar \sim$}}\raise 1.0pt \hbox {$<$}}

\title{Bound states of multi-nucleon channels in $N_f=2+1$ lattice QCD}

\ShortTitle{Bound states of multi-nucleon channels in $N_f=2+1$ lattice QCD}

\author{\speaker{Takeshi~Yamazaki}$^{ab}$,
Ken-ichi~Ishikawa$^{cb}$, Yoshinobu~Kuramashi$^{deb}$ and Akira~Ukawa$^{d}$
\\ \\
$^a$Kobayashi-Maskawa Institute for the Origin
 of Particles and the Universe, Nagoya University, Naogya, Aichi 464-8602, 
Japan\\
$^b$RIKEN Advanced Institute for Computational Science,
Kobe, Hyogo 650-0047, Japan\\
$^c$Department of Physics,
Hiroshima University,
Higashi-Hiroshima, Hiroshima 739-8526, Japan\\
$^d$Center for Computational Sciences, University of Tsukuba,
Tsukuba, Ibaraki 305-8577, Japan\\
$^e$Graduate School of Pure and Applied Sciences,
University of Tsukuba, Tsukuba, Ibaraki 305-8571, Japan\\
}

\abstract{We calculate the energies for multi-nucleon ground states with
the nuclear mass number less than or equal to 4 
in 2+1 flavor QCD at the lattice spacing of $a = 0.09$ fm
employing a relatively heavy quark mass corresponding to $m_\pi = 0.51$ GeV.
We investigate the volume dependence of the energy shift 
of the ground state and the state of free nucleons 
to distinguish a bound state from attractive scattering states.
From the investigation we conclude that $^4$He, $^3$He,
deuteron and dineutron are bound at $m_\pi = 0.51$ GeV.
We compare their binding energies with those in our quenched studies
and also with some recent investigations.}

\FullConference{The 30 International Symposium on Lattice Field Theory - Lattice 2012,\\
		June 24-29, 2012\\
		Cairns, Australia}

\begin{document}

\section{Introduction}

Lattice QCD has a potential ability to quantitatively
understand the nature of nuclei,
whose characteristic feature is a hierarchical structure in the strong interaction. 
The nuclear binding energy is experimentally known to be
about 10 MeV per nucleon,  which is much smaller than the typical energy scale of hadrons.
A measurement of the binding energies is therefore 
the first step for direct investigation of nuclei in lattice QCD.  

In this direction we carried out a first attempt to measure   
the binding energies of the $^4$He and $^3$He nuclei in quenched QCD 
with a rather heavy quark mass corresponding to $m_\pi=0.80$ GeV, thereby 
avoiding heavy computational cost~\cite{Yamazaki:2009ua}. 
We followed this work by a renewed investigation of the bound state for 
the two-nucleon channel 
in quenched QCD at the same quark mass, which found that not only the deuteron 
in the $^3$S$_1$ channel 
but also the dineutron in the $^1$S$_0$ channel is bound~\cite{Yamazaki:2011nd}.
However, the bound state in the $^1$S$_0$ channel has 
not been observed in the experiment.

The situation of the two-nucleon channel is little complicated.
There are two approaches to study the two-nucleon channel, such as
direct calculation of the ground state energy~\cite{Fukugita:1994na,Fukugita:1994ve,Beane:2006mx,Beane:2009py,Beane:2011iw,Beane:2012vq} and 
indirect calculation using the effective potential~\cite{Ishii:2006ec}.
The studies with former approach, which is same as in our previous work, reported 
a possibility of the bound state in both the channels
at $m_\pi=0.39$ GeV in 2+1 flavor QCD~\cite{Beane:2011iw},
and the bound state formations at $m_\pi=0.81$ GeV in 
3-flavor QCD~\cite{Beane:2012vq}.
The pion mass of the 3-flavor case is comparable to the one of
our quenched case, but the binding energies in the 3-flavor
are about twice larger than ours.
To understand the situation of the two-nucleon channel, we need
furthre investigations with less systematic errors.

In order to reduce systematic errors,
such as the quenched effect and heavier quark mass,
we extend our previous works to the dynamical quark calculation
with a lighter mass.
In this report we present our results of the binding energies 
of the helium nuclei, the deuteron and the dineutron on 2+1 flavor QCD 
with the degenerate $u,d$ quark mass corresponding to $m_\pi=0.51$ GeV. 
The details of this work have been already published in 
Ref.~\cite{Yamazaki:2012hi}.

\section{Simulation details}
\label{sec:details}

We generate 2+1 flavor gauge configurations with the 
Iwasaki gauge action~\cite{Iwasaki:2011jk} and 
the non-perturbative $O(a)$-improved Wilson quark action
at $\beta = 1.90$ with $c_{\rm SW} = 1.715$~\cite{Aoki:2005et}.
The lattice spacing is $a=0.8995(40)$ fm, corresponding to $a^{-1} = 2.194(10)$ GeV,
determined with $m_\Omega=1.6725$ GeV~\cite{Aoki:2009ix}.
We take four lattice sizes, 
$L^3\times T = 32^3 \times 48$, $40^3 \times 48$, $48^3 \times 48$ 
and $64^3 \times 64$, 
to investigate the spatial volume dependence of the ground state energy 
shift between the multi-nucleon system and the free nucleons.
The physical spatial extents are 2.9, 3.6, 4.3 and 5.8 fm, respectively.
From the investigation we distinguish a bound state from
an attractive scattering state~\cite{Luscher:1986pf,Beane:2007qr,Beane:2003da,Sasaki:2006jn}.
Since it becomes harder to obtain a good signal-to-noise ratio at lighter quark masses for multi-nucleon systems~\cite{Lepage:1989hd,Fukugita:1994ve},
we employ heavier $u,d$ quark mass corresponding to 
$m_\pi = 0.51$ GeV and $m_N = 1.32$ GeV. On the other hand, 
the strange quark mass is close to the physical value.
The hopping parameters are $(\kappa_{ud},\kappa_s) = (0.1373316,0.1367526)$
which are chosen based on the previous results for $m_\pi$ and
$m_s$ obtained by PACS-CS Collaboration~\cite{Aoki:2008sm,Aoki:2009ix}.

We extract the ground state energies of the multi-nucleon systems 
and the nucleon state from the correlation functions
$\displaystyle{
G_{\mathcal{O}}(t) = \langle 0 | \mathcal{O}(t)
\overline{\mathcal{O}}(0) | 0 \rangle
}$
with $\mathcal{O}$ being appropriate operators
for $^4$He, $^3$He, two-nucleon $^3$S$_1$ and $^1$S$_0$ channels,  
and the nucleon state $N$.

We are interested in the energy shift between the ground state of the multi-nucleon system and the free nucleons on an $L^3$ box,
\begin{equation}
\Delta E_L=E_{\mathcal{O}}-N_N m_N
\label{eq:delE_L}
\end{equation}
with $E_{\mathcal{O}}$ being the lowest energy level for the multi-nucleon
channel, $N_N$ the number of nucleon and $m_N$ the nucleon mass. 
This quantity is directly extracted from the ratio of the multi-nucleon 
correlation function
divided by the $N_N$-th power of the nucleon correlation function
\begin{equation}
R(t) = \frac{G_{\mathcal{O}}(t)}{\left(G_N(t)\right)^{N_N}},
\label{eq:R}
\end{equation}
when $t$ is large enough.
The same source operator is chosen for the numerator
and the denominator.
We also define the effective energy shift as
\begin{equation}
\Delta E_L^{\mathrm{eff}} = \ln \left(\frac{R(t)}{R(t+1)}\right),
\label{eq:eff_delE_L}
\end{equation}
from which we check the plateau region.

\section{Results}
\label{sec:results}

\subsection{$^4$He nucleus}
\label{sec:4N}

\begin{figure}[!t]
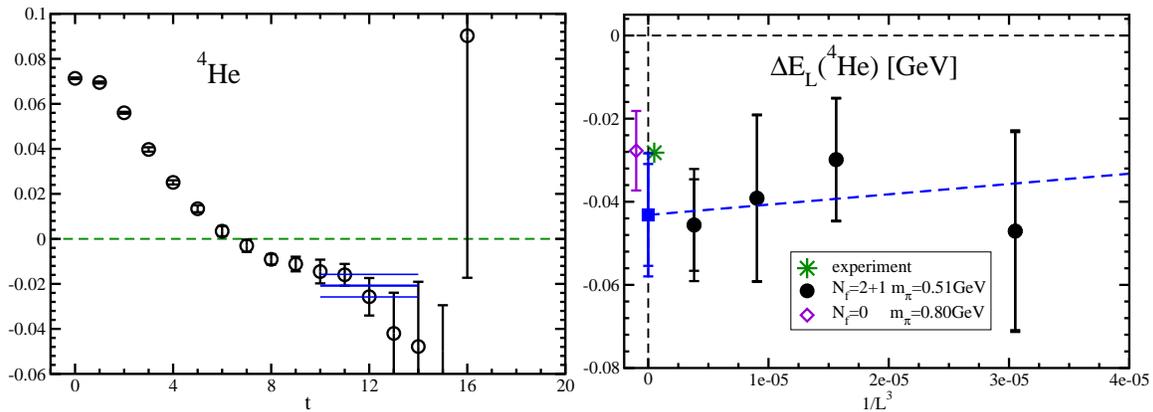

\includegraphics*[angle=0,width=0.5\textwidth]{fig02.eps}
\includegraphics*[angle=0,width=0.5\textwidth]{fig03.eps}
\caption{
The left panel shows effective energy shift 
$\Delta E_L^{\mathrm{eff}}$
for $^4$He channel on (5.8 fm)$^3$ box in lattice units.
Fit result with one standard deviation error band is expressed
by solid lines.
The right panel shows 
spatial volume dependence of $\Delta E_L$ in GeV units.
Outer bar denotes the combined error of statistical and 
systematic ones added in quadrature. Inner bar is for
the statistical error.
Extrapolated result in the infinite spatial volume limit is
shown by filled square symbol together with the fit line (dashed).
Experimental value (star) and quenched result (open diamond) 
are also presented.
\label{fig:h4}
}
\end{figure}

The effective energy shift $\Delta E^{\mathrm{eff}}_L$ defined in
Eq.~(\ref{eq:eff_delE_L}) is plotted in the left panel of 
Fig.~\ref{fig:h4}.
The signal is clear up to $t=12$, 
beyond which the statistical error increases rapidly.
The energy shift $\Delta E_L$ is extracted from 
$R(t)$ of Eq.~(\ref{eq:R}) by an exponential fit.
The fit result is denoted by the solid lines with the statistical error
band in the figure.
Results with similar quality are obtained on other volumes.

The right panel of Fig.~\ref{fig:h4} 
shows the volume dependence of $\Delta E_L$
as a function of $1/L^3$.
The inner bar denotes the statistical error and the outer bar represents the statistical and 
systematic errors combined in quadrature.
The negative energy shifts are obtained in all the four volumes.
We extrapolate the results
to the infinite volume limit with a simple 
linear function of $1/L^3$,
\begin{equation}
\Delta E_L = \Delta E_\infty + \frac{C_L}{L^3}.
\label{eq:linear}
\end{equation}
The systematic error is estimated from the variation of
the results obtained by alternative
fits which contain a constant fit of the data 
and a fit of the data obtained with a different fit range in $t$,
where the minimum or maximum time slice is changed by $\pm 1$.

The non-zero negative value obtained for the infinite volume 
limit $\Delta E_\infty$ shown in the figure
leads us to conclude 
that the ground state is bound in this channel for the quark mass. 
The binding energy $-\Delta E_\infty=43(12)(8)$ MeV, where the first error is statistical and the second one is systematic, 
is consistent with the experimental result of 28.3 MeV and 
also with the previous quenched result at
$m_\pi = 0.80$ GeV~\cite{Yamazaki:2011nd},
although the error is still quite large. 

A recent work in 3-flavor QCD at $m_\pi=0.81$ GeV reported a value 
110(20)(15) MeV for the binding energy of $^4$He nucleus~\cite{Beane:2012vq}.  
This is about three times deeper than our value. 
Whether this difference can be attributed to 
the quark mass dependence in unquenched calculations
needs to be clarified in future. 

\subsection{$^3$He nucleus}
\label{sec:3N}

\begin{figure}[!t]
\includegraphics*[angle=0,width=0.5\textwidth]{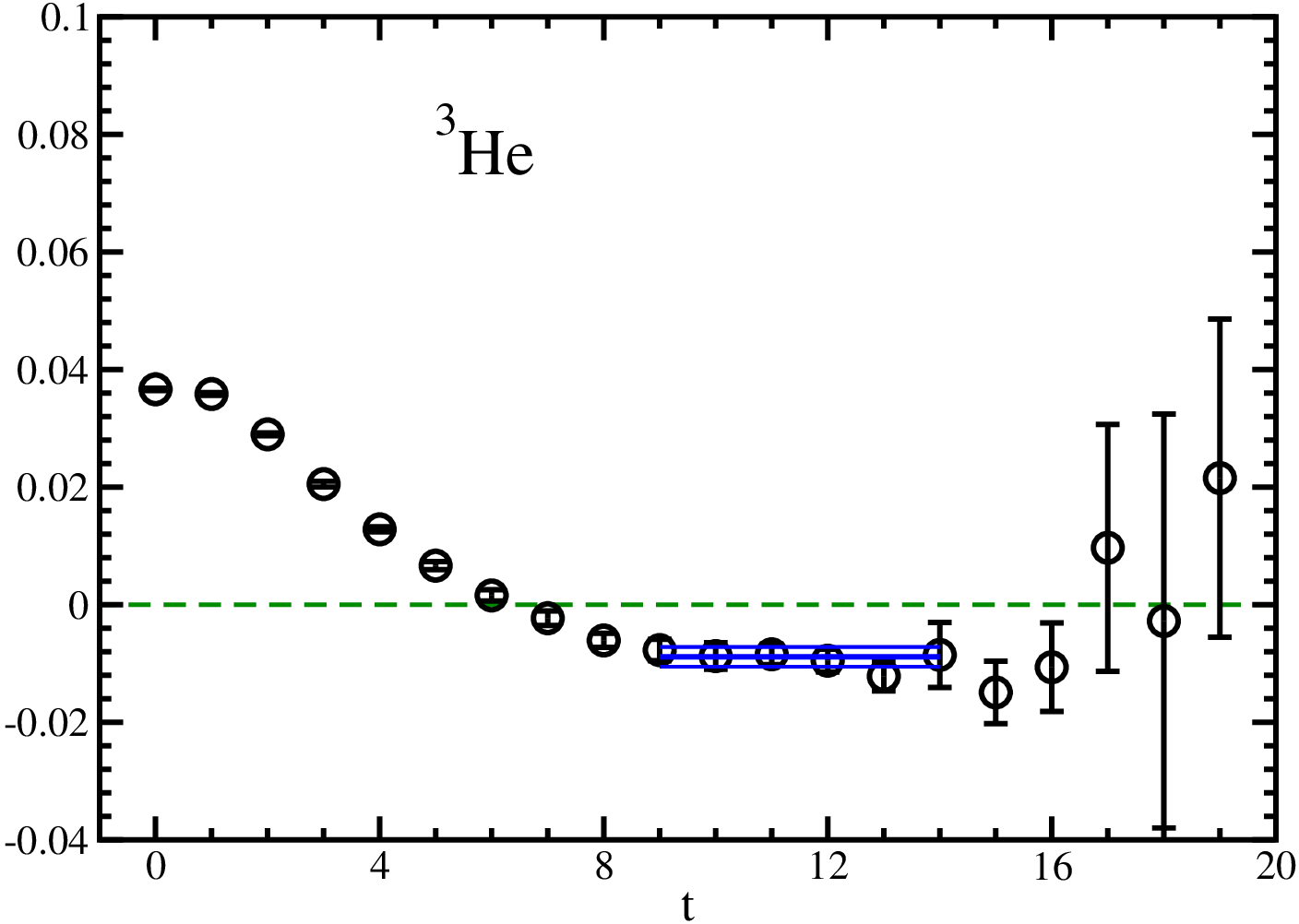}
\includegraphics*[angle=0,width=0.5\textwidth]{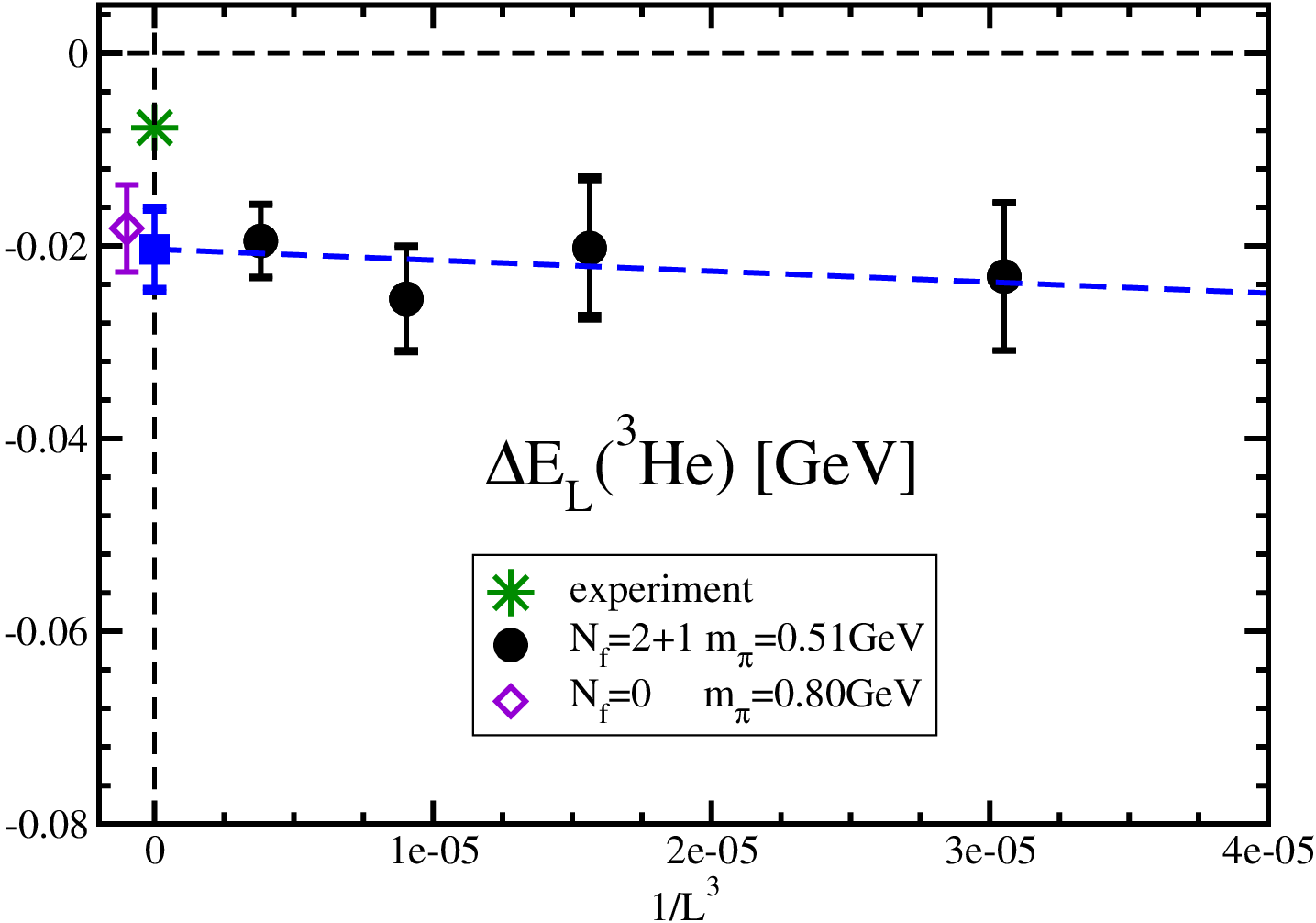}
\caption{
Same as Fig.~\protect\ref{fig:h4} for 
$^3$He channel.
\label{fig:h3}
}
\end{figure}

The left panel of Fig.~\ref{fig:h3} shows the effective energy shift 
$\Delta E^{\mathrm{eff}}_L$ of Eq.~(\ref{eq:eff_delE_L}).
An exponential fit of $R(t)$ in Eq.~(\ref{eq:R})
yields a negative value, which is denoted by the solid lines 
with the statistical error band in the figure.

The volume dependence is illustrated in the right panel of 
Fig.~\ref{fig:h3} as a function
of $1/L^3$ with the inner and outer error bars 
as explained in the previous subsection. 
We carry out a linear extrapolation of Eq.~(\ref{eq:linear}).
The systematic error is estimated in the same way as in the $^4$He channel.
The right panel of Fig.~\ref{fig:h3} shows that the energy shift extrapolated to the infinite spatial volume limit is non-zero and negative.
This means that the ground state is a bound state in this channel.
The value of $-\Delta E_\infty=20.3(4.0)(2.0)$ MeV is 
roughly three times larger than
the experimental result, 7.72 MeV,
though consistent with our previous quenched result at
$m_\pi = 0.80$ GeV~\cite{Yamazaki:2011nd}.

In 3-flavor QCD $-\Delta E_\infty=71(6)(5)$ MeV was reported~\cite{Beane:2012vq} 
at a heavier quark mass corresponding to $m_\pi=0.81$ GeV.  
Here again future work is needed to see if a quark mass dependence explains the difference from the experiment.

\subsection{Two-nucleon channels}
\label{sec:2N}

\begin{figure}[!t]
\includegraphics*[angle=0,width=0.5\textwidth]{fig06.eps}
\includegraphics*[angle=0,width=0.5\textwidth]{fig08.eps}
\caption{
Same as Fig.~\protect\ref{fig:h4} for $^3$S$_1$ channel.
\label{fig:3S1}
}
\end{figure}

\begin{figure}[!t]
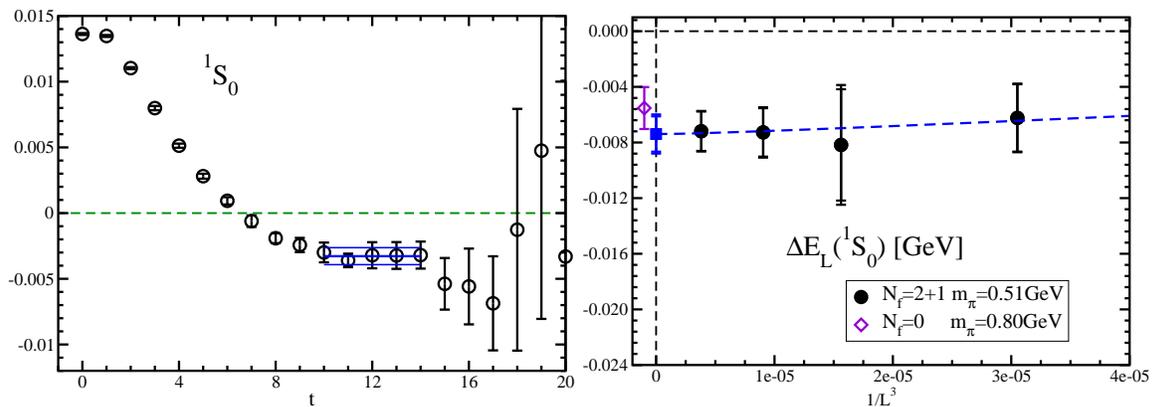

\includegraphics*[angle=0,width=0.5\textwidth]{fig07.eps}
\includegraphics*[angle=0,width=0.5\textwidth]{fig09.eps}
\caption{
Same as Fig.~\protect\ref{fig:h4} for $^1$S$_0$ channel.
There is no experimental value.
\label{fig:1S0}
}
\end{figure}

In the left panel of Fig.~\ref{fig:3S1} we show the 
time dependence for $\Delta E_L^{\mathrm{eff}}$ of Eq.~(\ref{eq:eff_delE_L})
in the $^3$S$_1$ channel.
The signals 
are lost beyond $t\approx 14$.
We observe negative values
beyond the error bars in the plateau region.
We extract the value of $\Delta E_L$ from an exponential fit for $R(t)$ of
Eq.~(\ref{eq:R}).

The left panel of Fig.~\ref{fig:1S0} 
shows the result for $\Delta E_L^{\mathrm{eff}}$ 
in the $^1$S$_0$ channel.
The value of $\Delta E_L^{\mathrm{eff}}$ 
is again negative beyond the error bars in the plateau region,
though the absolute value is smaller than in the $^3$S$_1$ case.
The energy shift $\Delta E_L$ is obtained in the same way 
as for the $^3$S$_1$ channel.

The volume dependences of $\Delta E_L$ 
in the $^3$S$_1$ and $^1$S$_0$ channels
are plotted as a function of $1/L^3$ in the right panel of 
Figs.~\ref{fig:3S1} and \ref{fig:1S0}, respectively.
There is little volume dependence for $\Delta E_L$, 
indicating a non-zero negative value in 
the infinite volume and a bound state, rather than the $1/L^3$ dependence 
expected for a scattering state, for the ground state for both channels.

The binding energies in the infinite spatial volume limit 
are obtained by fitting the data with a 
function including a finite volume effect on the 
two-particle bound state~\cite{Beane:2003da,Sasaki:2006jn},
\begin{equation}
\Delta E_L = -\frac{\gamma^2}{m_N}\left\{
1 + \frac{C_\gamma}{\gamma L} \sum^{\hspace{6mm}\prime}_{\vec{n}}
\frac{\exp(-\gamma L \sqrt{\vec{n}^2})}{\sqrt{\vec{n}^2}}
\right\},
\end{equation}
where $\gamma$ and $C_\gamma$ are free parameters, 
$\vec{n}$ is three-dimensional integer vector, 
and $\sum^\prime_{\vec{n}}$ denotes the summation without $|\vec{n}|=0$.
The binding energy $-\Delta E_\infty$ is determined from
\begin{equation}
-\Delta E_\infty = \frac{\gamma^2}{m_N} \approx 
2 m_N - 2\sqrt{m_N^2 - \gamma^2}.
\end{equation}
The systematic error is estimated from the variation 
of the fit results choosing different fit ranges
in the determination of $\Delta E_L$ and also using constant and linear fits 
as alternative fit forms.
We obtain the binding energies $-\Delta E_\infty$=11.5(1.1)(0.6) MeV 
and 7.4(1.3)(0.6) MeV for the $^3$S$_1$ and $^1$S$_0$ channels, respectively.
The result for the $^3$S$_1$ channel is roughly five times larger than
the experimental value, 2.22 MeV. 
Our finding of a bound state in the $^1$S$_0$ channel contradicts 
the experimental observation.
These features are consistent with our quenched results with a heavy
quark mass corresponding to $m_\pi=0.80$ GeV~\cite{Yamazaki:2011nd}.

The most recent study~\cite{Beane:2012vq} at a heavier quark mass of 
$m_\pi=0.81$ GeV in 3-flavor QCD found large values for the binding 
energies:
25(3)(2) MeV for the $^3$S$_1$ channel 
and 19(3)(1) MeV for the $^1$S$_0$ channel.
While all recent studies~\cite{Yamazaki:2011nd,Beane:2011iw,Beane:2012vq} are consistent with a bound ground state 
for both $^3$S$_1$ and $^1$S$_0$ channels when quark masses are heavy, 
quantitative details still need to be clarified.

\section{Conclusion and discussion}
\label{sec:summary}

We have calculated the binding energies for the helium nuclei, the
deuteron and the dineutron in 2+1 flavor QCD with $m_\pi=0.51$ GeV
and $m_N=1.32$ GeV. The bound states are distinguished from the attractive 
scattering states by investigating the spatial volume dependence of 
the energy shift $\Delta E_L$. 
While the binding energy for the 
$^4$He nucleus is comparable with the experimental value, those for the 
$^3$He nucleus and the deuteron are much larger than the experimental ones.
Furthermore we detect the bound state in the $^1$S$_0$ channel 
as in the previous study with quenched QCD, which is not observed in nature. 
To understand the discrepancy from the experimental results we need further
study of systematic errors in our results, 
especially for the heavier quark mass employed in the calculations.

\section*{Acknowledgments}
Numerical calculations for the present work have been carried out
on the HA8000 cluster system at Information Technology Center
of the University of Tokyo, on the PACS-CS computer 
under the ``Interdisciplinary Computational Science Program''
and the HA-PACS system
under HA-PACS Project for advanced interdisciplinary computational sciences
by exa-scale computing technology 
of Center for Computational Sciences in University of Tsukuba, 
on the T2K-Tsukuba cluster system at University of Tsukuba,
and on K computer at RIKEN Advanced Institute for Computational Science.
We thank our colleagues in the PACS-CS Collaboration for helpful
discussions and providing us the code used in this work.
This work is supported in part by Grants-in-Aid for Scientific Research
from the Ministry of Education, Culture, Sports, Science and Technology 
(Nos. 18104005, 18540250, 22244018) and 
Grants-in-Aid of the Japanese Ministry for Scientific Research on Innovative 
Areas (Nos. 20105002, 21105501, 23105708).

\bibliography{yamazaki}

\providecommand{\href}[2]{#2}\begingroup\raggedright\begin{thebibliography}{10}

\bibitem{Yamazaki:2009ua}
{\bf PACS-CS Collaboration} T.~Yamazaki, Y.~Kuramashi, and
  A.~Ukawa,  {\em Phys. Rev.}
  {\bf D81} (2010) 111504(R).

\bibitem{Yamazaki:2011nd}
T.~Yamazaki, Y.~Kuramashi, and A.~Ukawa,  
  {\em Phys. Rev.} {\bf D84} (2011) 054506.

\bibitem{Fukugita:1994na}
M.~Fukugita, Y.~Kuramashi, H.~Mino, M.~Okawa, and A.~Ukawa, 
  {\em Phys. Rev. Lett.} {\bf 73} (1994) 2176.

\bibitem{Fukugita:1994ve}
M.~Fukugita, Y.~Kuramashi, M.~Okawa, H.~Mino, and A.~Ukawa,
  {\em Phys. Rev.} {\bf D52} (1995) 3003.

\bibitem{Beane:2006mx}
S.~R. Beane, P.~F. Bedaque, K.~Orginos, and M.~J. Savage, 
  {\em Phys. Rev. Lett.} {\bf 97} (2006) 012001.

\bibitem{Beane:2009py}
{\bf NPLQCD Collaboration}, S. Beane {\em et.~al.}, 
  {\em Phys. Rev.} {\bf D81} (2010) 054505.

\bibitem{Beane:2011iw}
{\bf NPLQCD Collaboration}, S.~Beane {\em et.~al.},  
  {\em Phys. Rev.} {\bf D85} (2012) 054511.

\bibitem{Beane:2012vq}
{\bf NPLQCD Collaboration}, S.~Beane {\em et.~al.},  
  \href{http://xxx.lanl.gov/abs/1206.5219}{{\tt 1206.5219}}.

\bibitem{Ishii:2006ec}
N.~Ishii, S.~Aoki, and T.~Hatsuda,
  {\em Phys. Rev. Lett.} {\bf 99} (2007) 022001;
S.~Aoki, T.~Hatsuda, and N.~Ishii, 
  {\em Prog.Theor.Phys.} {\bf 123} (2010) 89.

\bibitem{Yamazaki:2012hi}
T.~Yamazaki, K.-i. Ishikawa, Y.~Kuramashi, and A.~Ukawa,
  {\em Phys. Rev.} {\bf D86} (2012) 074514.

\bibitem{Iwasaki:2011jk}
Y.~Iwasaki, 
  \href{http://xxx.lanl.gov/abs/1111.7054}{{\tt 1111.7054}}.

\bibitem{Aoki:2005et}
{\bf CP-PACS/JLQCD Collaborations}, S.~Aoki {\em et.~al.},
  {\em Phys. Rev.} {\bf D73} (2006) 034501.

\bibitem{Aoki:2009ix}
{\bf PACS-CS Collaboration}, S.~Aoki {\em et.~al.},
  {\em Phys. Rev.} {\bf D81} (2010) 074503.

\bibitem{Luscher:1986pf}
M.~L{\"u}scher, 
  {\em Commun. Math. Phys.} {\bf 105} (1986) 153.

\bibitem{Beane:2007qr}
S.~R. Beane, W.~Detmold, and M.~J. Savage,
  {\em Phys. Rev.} {\bf D76} (2007) 074507.

\bibitem{Beane:2003da}
S.~R. Beane, P.~F. Bedaque, A.~Parreno, and M.~J. Savage,
  {\em Phys. Lett.} {\bf B585} (2004) 106.

\bibitem{Sasaki:2006jn}
S.~Sasaki and T.~Yamazaki, 
  {\em Phys. Rev.} {\bf D74} (2006) 114507.

\bibitem{Lepage:1989hd}
G.~P. Lepage,
  \href{http://xxx.lanl.gov/abs/CLNS-89-971, C89-06-04}{{\tt CLNS-89-971,
  C89-06-04}}.

\bibitem{Aoki:2008sm}
{\bf PACS-CS Collaboration}, S.~Aoki {\em et.~al.},
  {\em Phys. Rev.} {\bf D79} (2009) 034503.

\end{thebibliography}\endgroup

\end{document}